\magnification = 1040
\baselineskip16pt
\abovedisplayskip 6pt plus2pt minus5pt
\belowdisplayskip 6pt plus2pt minus5pt
\vsize=9.7 true in \raggedbottom
\hsize=5.9 in
\hoffset= 4 true mm
\voffset= -0.1 true in
\font\bmit=cmmib10 \textfont9=\bmit \def\bmit{\fam9 }
\font\bx=cmbsy10
\textfont10=\bx \def\bx{\fam10 }

\mathchardef\pi="7119
\mathchardef\sigma="711B
\mathchardef\mu="7116
\mathchardef\nabla="7272
\parskip=5 pt
\interlinepenalty=50
\def\boxit#1{\vbox{\hrule\hbox{\vrule\kern1pc
   \vbox{\kern1pc#1\kern1pc}\kern1pc\vrule}\hrule}}
\def\sqr#1#2{{\vcenter{\vbox{\hrule height.#2pt
   \hbox{\vrule width.#2pt height#1pt  \kern#1pt
      \vrule width.#2pt}
     \hrule height.#2pt }}}}
\def\square{\mathchoice\sqr64\sqr64\sqr{4.2}3\sqr33}

\noindent{\bf Alternative potentials for the electromagnetic field  }

Shaun N. Mosley, Alumni, University of Nottingham NG7 2RD , England

e-mail: shaun.mosley@alumni.nottingham.ac.uk
\vskip 0.1 in

\beginsection Abstract

The electromagnetic field can be expressed in terms of two complex
potentials $ \alpha , \beta \, ,$ which are related to the Debye potentials.
The evolution equations for $ \alpha , \beta \, $ are derived, which are separable
either in parabolic coordinates (leading to the radiation fields) or in radial coordinates
(multipole fields). Potentials corresponding to focused wave fields as well as
plane waves are discussed. A conserved radiation density can be
constructed in terms of these potentials, which is positive (negative)
for positive (negative) helicity radiation.

\noindent {\bf PACS nos. 4110  } classical electromagnetism \hfil\break
\indent \qquad \qquad {\bf 4110H  } electromagnetic waves - theory

\vskip 0.1 in

\beginsection I. INTRODUCTION

The source-free electromagnetic field
has only two degrees of freedom per
space-time point. Some economy is achieved by reducing the six components of the
electric and magnetic fields $ ({\bf E} , {\bf B}) $ to the usual
4 potentials $ (\Phi , {\bf A}) $ which satisfy the covariant Lorentz
condition $ \partial_t \Phi + {\bx \nabla} \cdot {\bf A} = 0 \, ,$ or the
$ \Phi $ component may be dispensed with provided that the three potentials
$ {\bf A} $ satisfy the non-covariant Coulomb
gauge $ {\bx \nabla} \cdot {\bf A} = 0 \, .$ The
most economical
way to express the two degrees of freedom of the free electromagnetic field
is in terms of the two real Debye potentials$^{1,2} $  $
(\psi_1 , \psi_2 ) \, .$ Here we introduce a pair of complex potentials
$ \{ \alpha , \beta \} \, $ from which the electromasgnetic field can be derived
via (8) below. Some advantages of using these potentials are: a conserved
radiation density representing the difference of positive and negative
helicity radiation can be constructed, they accomodate singularities in the field
(charged particles), and interesting focused wave solutions arise naturally
when solving the evolution equations (equation (A7)below).

We first review the Debye potentials
to which $ \{ \alpha , \beta \} \, $ are related. The electric and magnetic fields
are expressed in
terms of the Debye potentials by
$$
\left[ \, \matrix{{\bf E} \cr {\bf B} \cr } \right]
= \left[ \, \matrix{  - ({\bf x} \times {\bx \nabla}) \partial_t \cr
 {\bx \nabla} \times ({\bf x} \times {\bx \nabla})  } \right] \psi_1
+ \left[ \, \matrix{ - {\bx \nabla} \times ({\bf x} \times
{\bx \nabla})  \cr
- ({\bf x} \times {\bx \nabla}) \partial_t  \cr  } \right] \psi_2
\eqno (1) $$
where $ \psi \, $ are solutions of the wave equation
$$ ( \partial_t^2 \, - \nabla^2 ) \,\psi = 0 \, . \eqno (2)  $$
The standard text-books may not refer to the Debye potentials by that name$^3 ,$
but use either $ \psi \, $ or $ L^2 \psi \, $ in their
analysis of multipole radiation fields, where $ \psi $ are spherical
solutions of the wave equation - see for example
pp.432-433 of Reitz$^4 ,$ or
pp.745-746 of Jackson$^5 .$
The paper by Boukamp and Casimir$^6 $ discusses of the various essentially
equivalent approaches to multipole radiation, while Refs.\ 2,7 show that any
field $ ({\bf E} , {\bf B}) $ outside the source region can be expressed in
terms of the potentials $ \psi_1 , \psi_2  \, .$

We express (1) more concisely by defining the complex field
$ {\bf F} \equiv {\bf E} + i {\bf B} \, ,$ then
$$ \eqalignno{
{\bf F} &= [ - ({\bf x} \times {\bx \nabla}) \partial_t + i {\bx \nabla}
\times
({\bf x} \times {\bx \nabla})  ] \; ( \psi_1 + i \psi_2 ) \cr
{\bf F} &= [ i \partial_t +  {\bx \nabla} \times \,  ]
\; {\bf L} \psi \, & (3) \cr  } $$
where $ {\bf L} \equiv - i {\bf r} \times {\bx \nabla} \, $ is the
angular momentum operator and
$ \psi \equiv  - ( \psi_1 + i \psi_2 ) \, .$
Conversely
$$ {\bf r} \cdot {\bf F} = {\bf r} \cdot {\bx \nabla} \times {\bf L} \psi
= i L^2 \psi  \eqno (4) $$
and the operator $ L^2 $ is, in principle, invertible.$^8 $  The source-free
Maxwell's
equations are
$$
[ i \, \partial_t - {\bx \nabla} \times \, ] \, {\bf F}  = 0 \qquad
{\bx \nabla} \cdot  {\bf F}  = 0 \, . \eqno (5a,b) $$
It is readily verified that substituting (3) into the above yields an
identity, given (2).

Despite the fact that the two Debye potentials (or the single complex
Debye potential $ \psi $~) precisely contain the two degrees of freedom of the
electromagnetic field, they are seldom used outside magnetostatics and multipole
field analysis. Some  disadvantages
spring to mind: \hfil\break
\noindent (a) the field $ {\bf F} \equiv {\bf E} + i {\bf B} \, $ is
expressed by second order differential operators acting on the potentials, so
then for example the usual energy density of the field $ ({\bf F}^* \, {\bf F}
) $ is a cumbersome expression in terms of $ \psi \, ,$ \hfil\break
\noindent (b) there is no regular closed form potential $ \psi \, $ representing
a plane wave field. The standard texts$^9 $ express the plane
wave  potential as an infinite sum over $
l $ of the spherical harmonics $
Y_{l,\pm 1} \, ,$  although this actually represents plane waves outgoing in
each direction from the $ \, (\theta = \pi / 2 ) \, $ plane, \hfil\break
\noindent (c) the Debye potential formalism is ill suited to coping with
point singularities representing charged particles in the field, i.e. there is
no regular potential $ \psi $ representing the monopole field
at the origin, as can be seen immediately from (4). See Ref.1 for irregular
potentials $ \psi $ representing
the monopole field. \hfil\break
\noindent We will now show that,
given (3), the field $ {\bf F} \, $
can be expressed in terms of two complex potentials $ \alpha , \beta \, $ acted
on by first order differential operators.  These potentials $
( \alpha , \beta ) \, $ admit charged
particle solutions, for static and moving charges, and the potentials
representing the plane wave fields are fairly simple.

From (3) we calculate the field $ ({\bf F} + i {\bf \hat{x}} \times {\bf F} ) \,
$ in terms of $
\psi \, ,$ where $ {\bf \hat{x}} \equiv {{\bf x} \over r } \,
.$  We use the operator identities$^2 $
$$ {\bx \nabla} \times {\bf L}
= ( - i {\bf x} \nabla^2 + i {\bx \nabla} ( \partial_r r ) \, )\, ,
\; \; {\bf \hat{x}} \times {\bf L} = i \, ( {\bx \nabla} r
- \partial_r {\bf x} ) \, , \; \;
{\bf \hat{x}} \times ( {\bx \nabla} \times {\bf L} )
= - {\bf L} ( \partial_r r ) \, . $$
Then
$$ \eqalignno{
{\bf F} + i {\bf \hat{x}} \times {\bf F}
&= [  i \, \partial_t {\bf L} - i {\bf x} \nabla^2 +
i {\bx \nabla}
( \partial_r r )  ] \; \psi \,
+ \, i \, [ \, - ( {\bx \nabla} r
- \partial_r {\bf x} ) \partial_t  - {\bf L} ( \partial_r r )  ] \; \psi \cr
{\bf F} + i {\bf \hat{x}} \times {\bf F}
&= [ {\bx \nabla} + {\bf \hat{x}} \partial_t - {1 \over r} {\bf L} ]
\, \alpha  & (6a) \cr
\noalign{\noindent where }
\alpha &= i ( \partial_r r \psi - \partial_t r \psi ) \, . & (7a) \cr
\noalign{\noindent To derive (6a) we have used (2). Similarly  }
{\bf F} - i {\bf \hat{x}} \times {\bf F}
&=
[ {\bx \nabla}
- {\bf \hat{x}} \partial_t + {1 \over r} {\bf L} ]
\, \beta & (6b)  \, \cr
\noalign{\noindent with }
\beta &= i ( \partial_r r \psi + \partial_t r
\psi ) \, . & (7b) \cr
} $$
Note that given $ ({\bf F} + i {\bf \hat{x}} \times {\bf F} ) \, $
there is no straightforward algebraic
manipulation to derive $ {\bf F} \, $ due to the fact that
the matrix $ ( I + i {\bf \hat{x}} \times ) $ is singular. We add (6a,b)
obtaining
$$ 2 {\bf F} = [ {\bx \nabla} + {\bf \hat{x}} \partial_t - {1
\over r} {\bf L}
] \, \alpha  + [ {\bx \nabla}
- {\bf \hat{x}} \partial_t + {1 \over r} {\bf L} ] \, \beta \, ,
 \eqno (8) $$
which is the basic formula expressing $ {\bf F} $ in terms of
$ \{\alpha , \beta \} \, .$
Dotting (6a,b) with $ {\bf \hat{x}} \, $ yields the constraint
$$ (\partial_r + \partial_t ) \alpha =  F_r = (\partial_r - \partial_t )
\beta \,  \eqno (9) $$
where $ F_r \equiv {\bf \hat{x}} \cdot {\bf F} \, .$
The operators $ ( \partial_r + \partial_t ) \, , \,
( \partial_r -
\partial_t ) \, $ are the radial differential
operators on the future, past lightcones respectively. Given the radial
field $ F_r (t,{\bf x}) \, ,$ then (9) enables us to calculate $ \alpha , \beta
\, $ as follows. To carry out the integration on the future (past) lightcone we
introduce the advanced (retarded) times $ S $ and $ T $ defined by
$$ S =  t - r \, , \qquad  T  =  t + r \, . \eqno (10) $$
The hypersurfaces $ S = S_0 , S_1... \, $
are future lightcones centered at the origin,
while the
hypersurfaces $ T = T_0 , T_1... \, $ are past lightcones. Then
$ F_r (S) \, $ or $ F_r (T) \, $ are obtained by substituting $ t = S + r \, $
or $ t = T - r \, $ into the expression for $ F_r (t) \, .$ A solution
of (9) is then
$$ \alpha (S) = \int_0^{r } dr \; F_r (S) \, \qquad
\beta (T) = \int_0^{r } dr \; F_r (T) \, . \eqno (11) $$
Maxwell's equations can alternatively be formulated from the start using
advanced or retarded time,$^{10} $ then $ \alpha \, $ or $ \beta \, $ is found
from $ F_r
\, $ by a straightforward radial integration.

We now find the evolution equations for $ {\alpha } , \beta \, ,$ noting that
$ \alpha = i ( \partial_r r - \partial_t r ) \psi \, $
does not satisfy the wave equation like $ \psi \, ,$ because the operator
$ ( \partial_r r - \partial_t r ) \, $ does not commute with the
d'Alembertian operator $ \square \equiv ( \partial_t^2 - \nabla^2 ) \, .$
The d'Alembertian may be expressed as follows,
$$
( \partial_t^2 - \nabla^2 ) \psi
= [ - {1 \over r}
(\partial_r + \partial_t ) (\partial_r - \partial_t ) r
\, + \, {1 \over r^2} L^2 ] \, \psi = 0 \, . \eqno (12) $$
Now multiply (11) by $ {i \over r^2}
(\partial_r - \partial_t ) r^3 $ obtaining
$$ \eqalignno{
 [ - {1 \over r^2}(\partial_r &- \partial_t) r^2 (\partial_r + \partial_t) \,
+ \, {1 \over r^2} L^2 ] i (\partial_r - \partial_t ) r \psi \cr
=& [ \partial_t^2 - \nabla^2
- \, {2 \over r} \partial_t ] {\alpha } = 0 \, . & (13a) \cr
\noalign{\noindent so that the evolution operator for $
\alpha \, $ is a
modified d'Alembertian, having the extra term $ - {2 \over r} \partial_t \, .$
Similarly }
& [ \partial_t^2 - \nabla^2
+ \, {2 \over r} \partial_t ] \beta = 0 \, . & (13b) \cr
}  $$

We see from (8) that adding any function $ f ( t - r ) $ to $ \alpha \, ,$  or
adding any function $ g ( t + r ) $ to $ \beta \, $ leaves $ {\bf F} \, $
unchanged.  Also
the evolution operators (13) for $ \alpha , \beta , $ annihilate $ f ( t - r )
, g ( t + r ) \, $ respectively.
So the gauge
transformations for $ \alpha , \beta \, $ are
$$ \eqalign{
\alpha &\rightarrow \alpha' \, , \qquad \qquad
\alpha' = \alpha + f ( t - r ) \cr
\beta &\rightarrow \beta' \, , \qquad \qquad
\beta' = \alpha + g ( t + r ) \, . \cr
} \eqno (14) $$

\penalty - 10000

\noindent{\bf II. THE POTENTIALS $ \{\alpha , \beta \} $ CORRESPONDING TO
RADIATION \hfil\break FIELDS }

\noindent We first establish some symmetries between $ \alpha , \beta \, .$
Consider
the parity transformation $ {\cal P} \, : {\bf x} \rightarrow - {\bf x} $ and
the  time reversal transformation $ {\cal T} \, : t \rightarrow - t \, .$ Under
the combined transformatioon $ {\cal P} {\cal T} \, $ the field transforms
as$^{ 11} $
$$ {\cal P T} \, {\bf F} (t, {\bf x} ) = - {\bf F} (- t, - {\bf x} )
\, .  $$
For brevity we will write a bar over any variable signifying its
$ {\cal P T} \, $ transform, e.g.\ $ {\bf \bar F} = {\cal P T} \, {\bf F} \;
, \, \bar\alpha = {\cal P T} \, \alpha \, $ etc. Both $ \alpha , \beta \, $
are scalars under $ {\cal P T} \, ,$ so that $ \bar\alpha (t, {\bf x} )
= \alpha ( - t, - {\bf x}
) \, .$
Maxwell's equations are invariant under the
$ {\cal P T} \, $ transformation, so that if $ {\bf F} $ satisfies
Maxwell's equations then so does $ {\bf F'} = {\bf \bar F} \, .$
In some cases of interest the field is even or odd under $ {\cal P T} \, ,$
by which we mean $ {\bf \bar F}_{ev} = {\bf F}_{ev} \, $ and $  {\bf
\bar F}_{odd} = - {\bf F}_{odd} \, $ (from now on we will use the terms
even and odd meaning even or odd under $ {\cal P T} \, .)$ Recall the fields
$ ({\bf F} + i {\bf
\hat{x}} \times
{\bf F} ) \,  , \,  ({\bf F} - i {\bf \hat{x}} \times
{\bf F} ) \, $ which are derived from $ \alpha , \beta \, $ respectively.
Then if the field is even/odd,
$$ ({\bf F}_{ev\atop odd} - i {\bf \hat{x}} \times {\bf F}_{ev\atop odd} ) =
\pm  {\cal P T} \,  ({\bf F}_{ev\atop odd} + i {\bf \hat{x}} \times {\bf
F}_{ev\atop odd} ) \, . \eqno (15) $$
In either of these cases the field is effectively given by one potential,
$ \alpha $ say, because after calculating $ ({\bf F} + i {\bf
\hat{x}} \times
{\bf F} ) \, $ from $ \alpha \, ,$ one can write down $ ({\bf F} - i {\bf
\hat{x}} \times {\bf F} ) \, $ using (15).
Applying the
$ {\cal P T} \, $ transformation to equations (6):
$$
{\bf \bar F} - i {\bf \hat{x}} \times {\bf \bar F}
= - [ {\bx \nabla} - {\bf \hat{x}} \partial_t + {1 \over r} {\bf L} ]
\, \bar\alpha  \qquad \qquad
{\bf \bar F} + i {\bf \hat{x}} \times {\bf \bar F}
= - [ {\bx \nabla} + {\bf \hat{x}} \partial_t - {1 \over r} {\bf L} ]
\, \bar\beta \, , $$
so that

$$ {\bf F} \rightarrow {\bf \bar F} \qquad \hbox{ is equivalent to } \qquad
\{ \alpha , \beta \} \rightarrow \{ - \bar\beta , - \bar\alpha \} \, .
\eqno (16) $$
From (16) we see that
$$  \big\{ {\bf F}_{ev} : \quad \beta = - \bar\alpha \, \big\}
\qquad \qquad  \big\{ {\bf F}_{odd} : \quad
\beta = + \bar\alpha \, \big\} \, . \eqno (17) $$

We now find the potentials $ \{\alpha \, , \, \beta \} $
corresponding to the radiation fields. Equations (11) enable us to calculate
the potentials for a
known field - which approach we follow in this section.
Alternatively we can look for solutions of the evolution equations for the
potentials (13), satisfying the constraint (9) - which approach we take in
appendix A. It turns out that these evolution equations (13) are separable
in either spherical or parabolic coordinates - the spherical solutions yield
the potentials corresponding to the multipole fields (which are singular at the
origin), while the parabolic solutions are potentials for
non-singular
fields, including the plane wave fields.

Consider the plane wave field $ {\bf F} \, $  propagating in the $ x_3 $
direction, the left circularly polarized (positive helicity) field $ {\bf F}_L
\, $ is
$$ {\bf F}_L =  \{ 1 , i , 0 \} \, e^{i k x_3 - i k t }
\qquad F_{rL} = ( \hat{x}_1 + i \hat{x}_2 ) \, e^{i k x_3 - i k t }
\,  \eqno (18) $$
and its $ {\cal P T} \, $ transform the
right circularly polarized (negative helicity) field $ {\bf F}_R \, $ is
$$ {\bf F}_R = {\bf \bar F}_L\,
= - \{ 1 , i , 0  \} \, e^{- (i k x_3 - i k t )}
\qquad F_{rR} = - ( \hat{x}_1 + i \hat{x}_2 ) \, e^{- (i k x_3 - i k t )}  \,
.$$
Then recalling (10) we integrate $ F_{rL} (S) =
( \hat{x}_1 + i \hat{x}_2 ) \, e^{ - i k S } e^{- i k r (1 - \hat{x}_3 ) } $ to
find $ \alpha \, ,$ and follow the equivalent procedure for $ \beta \, ,$
obtaining
$$ \eqalign{
\alpha_L &= \; \big( {\hat{x}_1 + i \hat{x}_2 \over 1 - \hat{x}_3 } \big)
  [ e^{ i k x_3 } - e^{ i k r } ] \, { e^{ - i k t } \,  \over i k}
\cr
\beta_L &=  - \big( {\hat{x}_1 + i \hat{x}_2 \over 1 + \hat{x}_3 } \big)
   [ e^{ i k x_3 } - e^{ - i k r } ] \,
{ e^{ - i k t } \,  \over i k} \, . \cr
} \eqno (19) $$
The second term in the square brackets does
not contribute to the field but its absence would make the potentials singular
due to the factors $ 1 / ( 1 \mp \hat{x}_3 ) ) \, $ in $ \alpha , \beta \, .$
Direct calculation verifies that
$$
[ {\bx \nabla} + {\bf \hat{x}} \partial_t - {1 \over r} {\bf L} ]
\, \alpha_L
= \big\{ \, 1 + \hat{x}_3 \, , \, i + i \hat{x}_3 \, , \,
- \hat{x}_1 - i \hat{x}_2  \, \big\}
\, e^{i k x_3 - i k t }    \,   $$
which is $ {\bf F}_L + i {\bf \hat{x}} \times {\bf F}_L \, ,$
noting that the operator $ [ {\bx \nabla}
+ {\bf \hat{x}} \partial_t - {1 \over r} {\bf L} ] $ commutes with the
$ ({x_1 + i x_2 \over r - x_3 } ) $ factor in $ \alpha_L \, ,$ and
annihilates the term $ e^{ i k (r - t) } \, .$
Similarly it can be verified that
$ [ {\bx \nabla} - {\bf \hat{x}} \partial_t + {1 \over r} {\bf L}
] \, \beta_L  = {\bf F}_L - i {\bf
\hat{x}} \times {\bf F}_L  \, .$

As $ {\bf F}_R  = {\bf \bar F}_L \, ,$ then recalling (16)
$$ \eqalign{
\alpha_R = - \bar\beta_L
&= - \big( {\hat{x}_1 + i \hat{x}_2 \over 1 - \hat{x}_3 } \big)
  [ e^{ - i k x_3 } - e^{ - i k r } ] \; { e^{ i k t } \,  \over i k} \cr
\beta_R = - \bar\alpha_L
&= \; \big({\hat{x}_1 + i \hat{x}_2 \over 1 + \hat{x}_3 } \big)
   [ e^{ - i k x_3 } - e^{ i k r } ] \;
{ e^{ i k t } \,  \over i k} \, . \cr } \eqno (20)  $$
The potentials for any plane wave propagating
in the $ x_3 $
direction are a linear superposition of $ \{ \alpha_L , \beta_L \} \, , \, \{
\alpha_R , \beta_R \} \, .$  Both $ \{ \alpha_L , \beta_L \} \, , \, \{ \alpha_R
, \beta_R \} \, $  are eigenstates of $  L_3 \, ,$
the angular momentum operator in the direction of propagation, with
eigenvalue $ + 1 \, .$

We obtain the potential $ \alpha_L ({\bf k}) $ for a left circularly polarized
wave propagating in the
$ {\bf \hat{k} } \equiv {\bf k} /|{\bf k} | \, $ direction with wavelength $ 2
\pi /
|{\bf k} |\, ,$  by first defining the null complex vector $ {\bf h} $
satisfying
$$ {\bf h} \times {\bf \hat{k} } = i {\bf h} \qquad
{\bf h}^* \cdot {\bf h} = 2 \, , \eqno (21) $$
which so defined is unique up to a phase factor.
Then
$$ \{ \alpha_L ({\bf k}) , \beta_L ({\bf k}) \}
=  \Big\{ - \big(
 { {\bf h} \cdot {\bf x}  \over r - {\bf \hat{k} }\cdot {\bf x} } \big)
  [ e^{ i {\bf k} \cdot {\bf x} } - e^{ i |{\bf k}| r } ] \; ,
\; \big({ {\bf h} \cdot {\bf x}  \over r + {\bf \hat{k}
}\cdot {\bf x} }
\big)
   [ e^{ i {\bf k} \cdot {\bf x} } - e^{ - i |{\bf k}| r } ] \, \Big\} \;
{ e^{ - i |{\bf k}| t } \,  \over i k } \, , \eqno  (22)  $$
and $ \{ \alpha_R ({\bf k}) , \beta_R ({\bf k}) \}
=  \{ - \beta_L ({\bf k}) , - \alpha_L ({\bf k}) \} \, .$
We check that $ \alpha_L ({\bf k}) \, , \, \beta_L ({\bf k}) $ are
eigenstates of $ ( {\bf \hat{k} } \cdot {\bf L} ) \, $ the angular momentum
operator in the direction of propagation, with
eigenvalue $ + 1 \, :$
$$ {\bf \hat{k} } \cdot
{\bf L} \, ({\bf h} \cdot {\bf x} )
= {\bf \hat{k} } \cdot ( - i {\bf x} \times {\bf h} )
= - i {\bf x} \cdot ( {\bf h} \times {\bf \hat{k} } )
= - i {\bf x} \cdot ( i {\bf h} )
= {\bf h} \cdot {\bf x} \, . \eqno (23)  $$
In the next section we investigate the orthonormality of the basis
potentials$ \{\alpha ({\bf k}) , \beta ({\bf k}) \}$ and
$ \, \{ \alpha ({\bf k'}) , \beta ({\bf k'}) \} \, .$

We next briefly consider the multipole radiation fields and their corresponding
potentials, noting
some symmetry relations under the $ {\cal P T} \, $
transformation. The Debye potential $ \psi \, $ in this case is a
spherical solution of the wave equation, then the potentials
$ \{\alpha \, , \, \beta \} $ follow from (7).
Consider
$$ \psi_{lm} = j_l (k r ) \;  Y_{lm} (\theta , \phi ) \;
\cos (k t ) \,  \eqno (24) $$
where  $ j_l $ is the spherical Bessel function of order $ l \, ,$ and $ Y_{lm}
$ are the spherical harmonics. This $ \psi_{lm} \, $ when substituted into (2)
yields $ {\bf F}_{lm}
\, ,$ the magnetic multipole field of order $(l,m) \, .$
(Multiplying the potential by $ i \, $ yields the corresponding electric
multpole field $ i \, {\bf F}_{lm} \, ).$ Then from (7)
the potentials $ \{ \alpha_{lm} , \beta_{lm} \} \, $
for the magnetic multipole field of order $(l,m) \, $ are
$$ \eqalign{
\alpha_{lm} &= i (\partial_r - \partial_t ) [  r j_l (k r ) \;
Y_{lm} (\theta , \phi ) \; \cos (k t ) ] \cr
\beta_{lm} &= i (\partial_r + \partial_t )  [ r j_l (k r ) \;
Y_{lm} (\theta ,
\phi ) \; \cos (k t ) ] \, .\cr } $$
We see that
$$ \beta_{lm} = (- 1 )^l \, \bar\alpha_{lm} $$
following the parity of $ Y_{lm} \, .$
This means, recalling (17), that $ {\bf F}_{lm} $ is odd or even, depending on
whether $ l \, $ is an even or odd number. (If we had considered the potential
$ \psi \, $ of (21) with a $ (\sin (k t ) )\, $ instead of a $ (\cos (k t )
) \, $ factor, then $ \beta'_{lm} = (- 1 )^{l+1} \, \bar\alpha'_{lm} \, .)$
Then, as discussed previously, $ {\bf F}_{lm} $ can be
calculated from either
one of the potentials $ \alpha , \beta \, .$

\vskip 0.2 in

\beginsection III. THE CONSERVED DENSITY $ \rho \, $ AND THE ENERGY
$ E \, $

\noindent {\bf A. The radiation density $ \rho \, .$ } \hfil\break
\noindent The electromagnetic field is the classical ``first-quantized''
version of the massless spin-1 photon field,
and as such should have a conserved density corresponding to the well-known
densities for the Klein-Gordon (spin-zero) or Dirac (spin-${1 \over 2} )
$
fields. The only text on classical (non-quantized) electromagnetism which
discusses a radiation density, or an inner product, for the electromagnetic
field that I have come across is the book by Good and Nelson,$^{12} $ otherwise
one has to turn to texts on QED such as that by Schweber.$^{13} $ For
an excellent recent review on this subject, see Ref.\ 14.
This inner product for the electromagnetic field involves
the non-local (integral) operator $ ({\nabla }^2 )^{- 1/2 } \, .$

The evolution
equations (7) for $ \alpha \, , \, \beta $  can be expressed in
the form
$$\eqalign{
&({\bx \nabla} - {\bf \hat{x}} \partial_t ) \cdot ({\bx \nabla} + {\bf \hat{x}}
\partial_t ) \alpha = 0 \cr
&({\bx \nabla} + {\bf \hat{x}} \partial_t ) \cdot ({\bx \nabla} - {\bf \hat{x}}
\partial_t ) \beta = 0 \, . \cr
}  \eqno (25) $$
It follows that
$$\eqalignno{
&({\bx \nabla} + {\bf \hat{x}} \partial_t ) \cdot
[ \alpha \, ({\bx \nabla} - {\bf \hat{x}} \partial_t ) \beta^*
- \alpha^* \, ({\bx \nabla} - {\bf
\hat{x}} \partial_t ) \beta
  ] \cr
& \qquad - ({\bx \nabla} - {\bf \hat{x}} \partial_t ) \cdot
[ \beta^* \, ({\bx \nabla} + {\bf \hat{x}} \partial_t ) \alpha
- \beta \, ({\bx \nabla} + {\bf \hat{x}} \partial_t ) \alpha^*
]  = 0 \, . & (26) \cr
}  $$
Collecting terms and multiplying by $ (- i ) \, $ we can write (4.3) in the form
of a continuity equation
$ {\partial_t} \rho  + {\bx \nabla} \cdot {\bf J} = 0 \, ,$  where
$$\eqalignno{
 \rho &= - { i \over 4 } \left[ \alpha \, (\partial_r -
\partial_t ) \beta^*
- \alpha^* \, (\partial_r - \partial_t ) \beta
+ \beta^* (\partial_r + \partial_t )  \alpha
- \beta \, (\partial_r + \partial_t ) \alpha^*
\right]  \cr
\rho &= - { i \over 4 }  ( \alpha - \beta )  \,
F_r^* \, \quad + \, \hbox{ CC } \quad
= - { i \over 4 } \left[ ( \alpha - \beta )  \, (\partial_r + \partial_t )
\alpha^* \,  \right] \quad + \, \hbox{ CC }
 & (27)  \cr
{\bf J} &= - { i \over 4 }  \left[
 \alpha \, ( {\bx \nabla} - {\bf \hat{x}} \partial_t )
\beta^*  \,  -
\, \alpha^* \,
( {\bx \nabla} - {\bf \hat{x}} \partial_t ) \beta  \,
- \, \beta^* \,
({\bx \nabla} + {\bf \hat{x}} \partial_t ) \alpha \,
+ \, \beta \, ( {\bx \nabla} + {\bf \hat{x}} \partial_t )
\alpha^*  \right] \,  \quad & (28)
\cr }  $$
where we have used (9), and  \hbox{ CC } stands for
the complex conjugate terms. Let us substitute the $ \{\alpha_L , \beta_L \} \,
$ of (19) into (27):
$$ \eqalignno{
\rho_L &= - { 1  \over 4 k } \left( - \big(
 {\hat{x}_1 + i \hat{x}_2 \over 1 -
\hat{x}_3 } \big)
  [ e^{ i k x_3 } - e^{ i k r } ] \,
- \, \big({\hat{x}_1 + i \hat{x}_2 \over 1 + \hat{x}_3 } \big)
   [ e^{ i k x_3 } - e^{ - i k r } ] \,  \right)  \;
\left( (\hat{x}_1 - i \hat{x}_2 ) \, e^{ - i k x_3 } \right)
+ \; \hbox{ CC } \cr
&= \, { 1  \over 2 k } \left(  \big(
 {1 + \hat{x}_3 } \big)
  [ 1 - \cos (k r -  k x_3 ) ] \,
+ \, \big({ 1 - \hat{x}_3 } \big)
   [ 1 - \cos ( k  r + k x_3 ) ] \,  \right)  \; \geq \, 0 \, . \cr
 } $$
Inserting $ \{\alpha_R , \beta_R \} = - \{
\bar\beta_R ,
\bar\alpha_R \} \, $ into (27), we obtain $ \rho_R = - \rho_L \, .$  Hence $
\rho_L \, $ is non-negative and $ \rho_R \, $ is non-positive.

The density $ \rho \, $ has close similarity with the Klein-Gordon density
for a spin-zero particle: $ \rho_{KG}
= - {i \over 2} (\phi^* \partial_t \phi - \phi \partial_t \phi^* ) \, .$ In this
case solutions $ \phi $ with time dependence $ e^{- i \omega t } $ with $
\omega $ positive are regarded as particle solutions, those with time
dependence $
e^{+ i \omega t } $ are regarded as anti-particle solutions.
Analagously we can regard the right circularly polarized (negative helicity)
field as the `anti-photons' of the left circularly polarized (positive
helicity) photon field. There appears to be no comparable
radiation density in the literature, although of course one can project out
the different polarizations when the field is expressed as a Fourier
integral.$^{15,16} $

We now investigate the orthogonality of the potentials for two
different
fields $ {\bf F}_1 $ and $ {\bf F}_2 \, .$ It will be convenient to
represent the potentials for the field $ {\bf F}_1 $ by the
$ 2 \times 1 $ matrix $ U_1 \equiv \left[ \matrix{\alpha_1 \cr \beta_1 \cr}
\right]
\, ,$  then we construct the following indefinite
scalar product space
$$ \langle {U}_1 | {U}_2 \rangle \equiv \int d^3 x \; \rho_{1 2}
\eqno (29) $$
where $ \rho_{1 2} \, $ is the conserved density
$$\eqalignno{
 \rho_{1 2}
&= - { i \over 4 } \left[ U^T_1 \; (I \partial_t +
\sigma^3 \partial_r ) U_2^*
\,   - \, {U_2^T}^* \; (I \partial_t + \sigma^3 \partial_r ) U_1 \right]  \cr
\rho_{1 2} &= - { i \over 4 } \left[ ( \alpha_1 - \beta_1 )  \, F_{r2}^* \,
- \, (\alpha_2^* - \beta_2^* ) \,
F_{r1 } \right] \, . & (30)  \cr }  $$
Here $ U^T = \left[ {\alpha \quad \beta } \right] \, $ is the
transpose of $ U \, ,$ and $ \sigma^3 $ is the Pauli matrix
$ \left[ \matrix{1 & \;0 \cr 0 &-1 \cr } \right] .$ Note that when
$ U_1 = U_2 = U \, ,$ then $ \rho_{1 2} = \rho \, .$ In
appendix
B we show that
$$ \eqalign{
\langle U_L ({\bf k}) | U_L ({\bf k'}) \rangle &= (2 \pi )^3
\, k \; \delta ({\bf k} - {\bf k'})  \cr
\langle U_R ({\bf k}) | U_R ({\bf k'}) \rangle
& = - (2 \pi )^3  \, k \;  \delta ({\bf k} - {\bf k'}) \cr
\langle U_L ({\bf k}) | U_R ({\bf k'}) \rangle &= 0 \, . \cr }
\eqno (31)  $$
The relations (31) enable one to project out the positive/negative helicity
states of momentum $ {\bf k} \, ,$ and show that  $ \rho \, $ integrated over
all space is the amount
of left minus right circularly polarized
radiation.

\vskip 0.1 in

\noindent {\bf B. The energy density  $ E \, $}
\hfil\break
\noindent As a consequence of Maxwell's equations (5)
$$ \partial_t ({1 \over 2} {\bf F} \cdot {\bf F}^* ) + {\bx \nabla} \cdot
( {i \over 2} \, {\bf F} \times {\bf F}^* ) = 0 \,   $$
where $ E = {1 \over 2} {\bf F} \cdot {\bf F}^* \equiv {1 \over 2}
( {\bf E}^2 + {\bf B}^2 ) \, $ is the usual energy density, and
$ {i \over 2} \, {\bf F} \times {\bf F}^*
\equiv {\bf E}
\times {\bf B} \, $ is the momentum density. To simplify
$  {\bf F} \cdot {\bf F}^* \, $ in terms of $ \alpha , \beta , $ we note the
following: that from (8) we can express $ {\bf F} \, $ as$^{17} $
$$ {\bf F} = {\bx \nabla}_{\uparrow} \alpha
+ {\bx \nabla}_{\downarrow} \beta + {\bf \hat{x}} F_r \eqno (32) $$
with $ F_r \, $ expressed in terms of  $ \alpha , \beta \, $ by (9), and
the operators $ {\bx \nabla}_{\uparrow} , {\bx \nabla}_{\downarrow} $
are defined as
$$ {\bx \nabla}_{\uparrow}
\equiv { 1 \over 2 } \, ( {\bx \nabla}
- {\bf \hat{x}} \, \partial_r + i {{\bf \hat{x}} \times {\bx \nabla} } ) \, ,
\qquad {\bx \nabla}_{\downarrow} = ({\bx \nabla}_{\uparrow} )^*
\equiv { 1 \over 2 } \, ( {\bx \nabla}
- {\bf \hat{x}} \, \partial_r - i {{\bf \hat{x}} \times {\bx \nabla} } ) \, .
\eqno (33) $$
These operators have the property that for any scalars $ \chi , \zeta $
$$ \eqalign{
&{\bx \nabla}_{\uparrow} \chi \cdot {\bx \nabla}_{\uparrow} \zeta
= {\bx \nabla}_{\downarrow} \chi \cdot
{\bx \nabla}_{\downarrow} \zeta
= {\bf \hat{x}} \cdot {\bx \nabla}_{\uparrow} \chi
= {\bf \hat{x}} \cdot {\bx \nabla}_{\downarrow} \chi = 0 \cr
&{\bx \nabla}_{\uparrow} \chi \cdot {\bx \nabla}_{\downarrow} \zeta
= { 1 \over 2 } [ ({ {\bf \hat{x}} \times {\bx \nabla} \chi }) \cdot
({ {\bf \hat{x}} \times {\bx \nabla} \zeta })
+ i {\bf \hat{x}} \cdot ({\bx \nabla} \chi \times {\bx \nabla} \zeta ) ] \, .
\cr } \eqno (34) $$
Then substituting in (32) into $ ({\bf F} \cdot {\bf F}^*) \, ,$ and with
the
aid of (34), we find
$$ \eqalignno{
2 \, E &= {\bf F} \cdot {\bf F}^*
= [{\bx \nabla}_{\uparrow} \alpha
+ {\bx \nabla}_{\downarrow} \beta + {\bf \hat{x}} F_r ] \cdot
[{\bx \nabla}_{\downarrow} \alpha^*
+ {\bx \nabla}_{\uparrow} \beta^* + {\bf \hat{x}} F_r^* ] \cr
2 \, E &= | {\bx \nabla}_{\uparrow} \alpha |^2  + | {\bx \nabla}_{\downarrow}
\beta |^2 + | F_r |^2 & (35) \cr } $$
with $ F_r $ given by (9). We
will further expand (35) as we shall see that $ E $ contains a divergence term,
which
can be removed such that the resulting energy $ E' $ is
still conserved. Again using (34),
$$ \eqalignno{
4 \, E &= |{ {\bf \hat{x}} \times {\bx \nabla} \alpha }|^2
+ |{ {\bf \hat{x}} \times {\bx \nabla} \beta }|^2 + 2 \,| F_r |^2
+ i {\bf \hat{x}} \cdot [({\bx \nabla} \alpha \times {\bx \nabla} \alpha^* )
- ({\bx \nabla} \beta \times {\bx \nabla} \beta^* )] \, ,  & (36)  \cr
} $$
and the last terms of (36) can be expressed as
$ i {\bx \nabla} \cdot [ - \alpha  \, ({\bf \hat{x}} \times {\bx
\nabla}
\alpha^* )  +  \beta \, ({\bf \hat{x}} \times {\bx \nabla} \beta^* ) ] \, ,$
so that the modified energy density $ E' $
$$ \eqalignno{
4 \, E' &= |{ {\bf \hat{x}} \times {\bx \nabla} \alpha }|^2
+ |{ {\bf \hat{x}} \times {\bx \nabla} \beta } |^2 + 2 \, | F_r |^2
 & (37)  \cr } $$
is also a conserved non-negative density, so is also a plausible candidate for
the energy density.

\vskip 0.2 in

\beginsection IV. THE POTENTIALS CORRESPONDING TO THE FIELD OF A
CHARGED PARTICLE

\noindent First consider the static case, then the constraint $ (\partial_r +
\partial_t ) \alpha = (\partial_r - \partial_t )
\beta \, $ implies $ \beta = \alpha \, ,$ then from (8) we obtain
$$ {\bf F}_{\rm stc} = {\bx \nabla} \alpha_{\rm stc} $$
so that
$$ Re[\alpha ] = - \Phi \quad \qquad Im[\alpha ] = \Phi_m $$
where $ \Phi \, $ is the usual scalar potential, and $ \Phi_m \, $ is the
magnetic scalar potential.
The potentials $ \Phi \, , \, \Phi_m \, $ for the
electrostatic,
magnetostatic fields are well known. The energy $ E' \, $ for the static field
is $ E' = {1 \over 2} |{ {\bx \nabla} \alpha }|^2 \, .$

Magnetic monopole fields are also accomodated:
the potential for a magnetic mono-pole at position $ {\bf a} \, $ is just
$ \alpha = \beta = i \, q_m / |{\bf x} - {\bf a} | \, .$
Note that if we use (11) to derive $ \alpha \, $ from the radial field
$$ F_r = q ( r - {\bf \hat{x}} \cdot {\bf a} ) / |{\bf x} - {\bf a} |^3 \, $$
of a charge at position
$ {\bf a} \, ,$ we obtain
$ \alpha = \int_0^{r } dr \; F_r = - ({q / |{\bf x} - {\bf a} | } )
+ ({q / | {\bf a} | }) \, .$
For non-radiation fields, we can lose the constant of integration by instead
obtaining $\alpha \,$ from $ F_r \, ,$ so that (9) is satisfied, as follows
$$ \alpha (S)
= \int_\infty^{r } dr \; F_r (S) \, . \eqno (38) $$
We will use (38) to determine the potentials $ \{ \alpha , \beta \} \, $ for a
uniformly moving charge. The field of a  charge passing through the origin in
the
$ x_1 \, $ direction is$^{18} $
$$ {\bf F} = \{ x_1 - v t , x_2 - i v x_3 , x_3 + i v x_2 \} \, {q \gamma \over
{r' }^{3 / 2} } \qquad F_r = {q \gamma ( r - v \hat{x}_1 t ) \over {r' }^{3 / 2}
} \eqno (39) $$
where $ {r' } = [ \gamma^2 (x_1 - v t )^2 + x_2^2 + x_3^2 ]^{1 / 2} \, .$
Substituting $ t = S + r \, $ into the expression for $ F_r \, ,$
$$ \eqalignno{
F_r (S) &= {q \gamma ( r - v \hat{x}_1 S - v \hat{x}_1 r  ) \over
[ \gamma^2 ( \hat{x}_1 r - v S - v r )^2 + (\hat{x}_2^2 + \hat{x}_3^2 )
r^2 ]
^{3 / 2}  }  & (40) \cr } $$
and integrating over $ r \, $ we obtain
$$ \eqalignno{
\alpha
&= \Big|_\infty^{r }
-  { q \gamma [ r (1 - \hat{x}_1 v ) + S ] \over
S [ \gamma^2 ( \hat{x}_1 r  - v S - v r )^2 + (\hat{x}_2^2 + \hat{x}_3^2 ) r^2
]^{1 / 2}  } \; \Big| \cr
\alpha&= - q \big({  t' - r' \over t - r } \big)  {  1 \over r' }
 \; & (41)  \cr} $$
where $ t' =  \gamma (t - v x_1 ) \, .$ This potential is non-singular except
at the charge position, because when $ t = r , \; r' = \gamma
(r - v x_1 ) \, ,$
and $ ({  t' - r' \over t - r } ) = \gamma \, .$ The field $ {\bf F} \, $ of
(39) is an even field (which is only the case because the particle is
passing through the origin), and so from (17)
$$ \beta = - \bar\alpha
= - q \big({  t' + r' \over t + r } \big)  {  1 \over r' } \, .
\qquad \qquad \qquad \qquad  $$

\vskip 0.05 in

\noindent {\bf V. OUTLOOK }

\noindent We have calculated the potentials $ \{\alpha , \beta \} $
corresponding to various fields: the simple
relation
(9) between the potentials and the radial field $ F_r \, $ makes the
calculation of the potential from a given field quite straightforward, or
one can solve the equations for the potentials from which one then calculates
the field, as in appendix A. Although
$ \{\alpha , \beta \} $ are scalars under rotations, they have complicated
transformation properties under the Lorentz transformations - we will discuss
these transformation properties elsewhere. (For the Lorentz transformation of
the Debye
potential $ \psi \, ,$ also complicated, see the paper by
Monroe.$^{19} )$
The stationary solutions for the potentials with time dependence $ e^{ - i k t
} \, $ $ ( e^{ + i k t
} \, ) $ correspond to the left (right) circularly polarized waves: the
orthogonality properties (31) suggest an alternative approach to second
quantizing the electromagnetic field, which we hope to address in the future.

Under the duality transformation $ ( {\bf E}
\rightarrow  {\bf B} \, , \; {\bf B} \rightarrow - {\bf E}
)\, ,$ or $ {\bf F}
\rightarrow  i \, {\bf F} \, ,$ the transformation of the potentials is just $
\{\alpha , \beta \} \rightarrow i \, \{\alpha , \beta \} \, .$  The absence of
any duality transformation for the usual potentials $ A^\mu
\equiv ( \Phi , {\bf A} ) $
has been commented on recently by Witten.$^{20} $  On the other hand the
interaction of a charged particle with the field is naturally described via
the potentials $ A^\mu \, ,$ by replacing the free momentum $ p^\mu \, $
with the gauge
invariant $ ( p^\mu - e A^\mu )\, ,$ whereas the role of the
$ \{\alpha , \beta \} $ potentials in gauge theory is not clear.

\vskip 0.2 in

\noindent {\bf Appendix A - Solutions of (13) in parabolic coordinates }

\noindent We solve the equations for the potentials (13). The spherical
solutions yield the multipole fields, here we discuss a few of the parabolic
solutions.
Defining  $ \lambda , \mu , \phi $
such that $ \lambda , \mu $  have the dimension of length:$^{21,22} $
$$ x_1 + i x_2 = 2
\sqrt{\lambda \mu } e^{i \phi } \, \qquad x_3 = \lambda -
\mu \qquad  r = \lambda + \mu \, ,$$
then$^{21,22} $
$$ \eqalignno{
{\bx \nabla }^2 &\equiv {1 \over \lambda + \mu } ( \partial_\lambda \lambda
\partial_\lambda +  \partial_\mu \mu \partial_\mu ) + {1 \over 4 \lambda \mu }
\partial_\phi^2 \cr
{\bx \nabla }^2 - \partial_t^2 + {2 \over r} \partial_t &\equiv
{1 \over
\lambda + \mu } ( \partial_\lambda \lambda  \partial_\lambda +  \partial_\mu
\mu \partial_\mu + 2 \partial_t )
+ {1 \over 4
\lambda \mu } \partial_\phi^2 - \partial_t^2 \cr
} $$
and with $ \alpha = f(\lambda ) g(\mu ) e^{ i m \phi } \, e^{ - i k t } $
then
$$ \eqalignno{
&[( \partial_\lambda \lambda  \partial_\lambda +  \partial_\mu
\mu \partial_\mu - 2 i k )
- {m^2 \over 4 \lambda } - {m^2 \over 4 \mu } + {k^2 \lambda }
+ {k^2 \mu }  ] \, f(\lambda ) g(\mu ) = 0 \, . \cr
} $$
Inserting a separation constant  $ {2 i k c } \, ,$ then
$$ \eqalignno{
&[( \partial_\lambda \lambda  \partial_\lambda - { i k (1 + 2 c)  }
- {m^2 \over 4 \lambda } + {k^2 \lambda } ] \,
f(\lambda ) = 0 & (A1) \cr
&[\partial_\mu \mu \partial_\mu - { i k (1 - 2 c) }
- {m^2 \over 4 \mu } + {k^2 \mu  } ] \, g(\mu ) = 0 \, . & (A2) \cr
} $$
The solution of (A1) is $ f(\lambda ) = \lambda^{m/2} e^{- i k \lambda }
{}_1 F_1 (1 + {\scriptstyle m \over 2} + c , m + 1 , 2 i k \lambda ) \, ,$
where $ {}_1 F_1 $ is the confluent hypergeometric function.
So the solution for $ \alpha $ is
$$ \alpha = (\lambda \mu )^{m/2} \, e^{ i m \phi } \,
e^{ - i k
(\lambda + \mu ) } \,  {}_1 F_1 (1 + {\scriptstyle{ m \over 2}} + c , m + 1 , 2
i k \lambda )  {}_1 F_1 (1 + {\scriptstyle{ m \over 2}} - c , m + 1 ,
2 i k \mu )  \, e^{ - i k t } \, . \eqno (A3) $$
The confluent hypergeometric function
reduces to a simpler expression for integer or half-integer values of $ c \, .$

\noindent {\bf The case $ m = 0 \, , \,  c = 0 \, .$} \hfil\break
\noindent As $ {}_1 F_1 (1 , 1  , i \zeta ) \equiv e^{ i \zeta } $ we have
$$ \eqalignno{
\alpha &=  \, e^{
- i k t } e^{ - i k ({\lambda + \mu}) } \,
e^{ 2 i k \lambda } e^{ 2 i k \mu }  =  \, e^{ - i k t }
e^{ i k ({\lambda + \mu}) } \,  \equiv  e^{ i k r - i k t }  \cr  }
$$
which potential substituted into (6a) yields a zero field.

\noindent {\bf The case $ m = 1 \, , \,  c = 1/2 \, .$} \hfil\break
\noindent Inserting these values of $ m , c $ into (A1), and using the
identities
${}_1 F_1 (2 \, , \, 2  \, , \, 2 i \zeta ) \equiv
\,  e^{2 i \zeta} \, , \hfil\break
{}_1 F_1 (1 \, , \, 2  \, , \, 2 i
\zeta ) \equiv {\sin {\zeta } \over
\zeta} \,  e^{i \zeta} \, ,$ then
$$ \eqalignno{
\alpha &= \sqrt{\lambda \mu } \,  e^{ i \phi } e^{- i k ({\lambda + \mu }) } \,
{}_1 F_1 ( 2 , 2 , 2 i k \lambda )
{}_1 F_1 ( 1 , 2 , 2 i k \mu ) \, e^{ - i k t } \cr
\alpha &= \sqrt{\lambda \mu } \,  e^{ i \phi } e^{- i k ({\lambda + \mu }) } \,
e^{2 i k \lambda }
\, {\sin ({k \mu }) \over k \mu } e^{i k \mu }
\, e^{ - i k t } \cr
\alpha &=  \sqrt{\lambda \mu } \, e^{ i \phi }
e^{ i k \lambda }  \, { \sin
({k \mu }) \over k \mu }  \, e^{ - i k t }
\,   \cr  } $$
which is the $ \alpha_L $ of (19).

\noindent {\bf The case $ m = 1 , c = 0 \, .$} \hfil\break
\noindent Substituting these values for $ m , c , $ into (3.2) we have
$$ \alpha_Z =  \sqrt{\lambda \mu } \, e^{ i m \phi } e^{ - i k
(\lambda + \mu ) }  {}_1 F_1 ({\scriptstyle{ 3 \over 2}} , 2  , 2 i k \lambda
)  {}_1 F_1 ({\scriptstyle{ 3 \over 2}} , 2  , 2 i k \mu )
 \, e^{ - i k t } \, . $$
As $ {}_1 F_1 ({\scriptstyle{ 3 \over 2}} , 2  , 2 i
\zeta ) \equiv
[ J_0 ({\zeta}) + i J_1 ({\zeta}) ] \, e^{i \zeta} \, ,$ where
$ J_0 \, , \, J_1 $ are the Bessel functions of order zero and one, we have
$$ \eqalignno{
\alpha_Z &=  \sqrt{\lambda \mu } \, e^{i \phi} [ (J_0 + i J_1 ) (k{\lambda}) ]
[ (J_0 + i J_1 ) (k{\mu}) ]  \, e^{ - i k t } & (A4) \cr
}  $$
where we have written $  (J_0 + i J_1 ) (k{\lambda}) $ as shorthand for
$ J_0 (k{\lambda}) + i J_1 (k{\lambda}) \, .$
For large $ \lambda \, ,$
$ \, \sqrt{\lambda } (J_0 + i J_1 )
(k{\lambda}) \simeq \sqrt{2 / \pi k } \,
e^{i k \lambda - \pi / 4} \, ,$
so that $ \alpha_Z $ is everywhere bounded. The potential $ \beta_Z $ such that
the constraint (9) is satisfied is
$$ \beta_Z = \sqrt{\lambda \mu } \, e^{i \phi} [ (J_0  - i J_1 ) (k{\lambda}) ]
[ (J_0  - i J_1 ) (k{\mu}) ]  \, e^{ - i k t } \, . \eqno (A5) $$
We can determine the
corresponding Debye potential $ \psi_Z \, ,$ such that $ i ( \partial_r r
\psi_Z - \partial_t r \psi_Z ) = \alpha_Z \, :$
$$ \psi_Z = {1 \over i k }
\left( { \sqrt{\lambda \mu } \over \lambda + \mu }
\right) \, e^{i \phi } [ J_0  (k{\lambda})  J_1 (k{\mu})  +  J_1 (k{\lambda})
J_0 (k{\mu}) ] \, . \eqno (A6) $$

The field $ {\bf F}_Z $ may be calculated by substituting (A4), (A5) into (8),
or (A6) into (3). After some labour one obtains the components of
$ {\bf F}_Z $
$$ \eqalignno{
2 \left[ \matrix{ F_1 \cr F_2 \cr F_3 \cr } \right]_Z
&= { e^{ - i k t } \over ( \lambda + \mu )  }
\left[ \pmatrix{ \lambda [ (J_0 + i J_1 ) (k{\lambda}) \,
 J_0
(k{\mu}) ] \, + \,  \mu [ J_0 (k{\lambda})  \,
 (J_0  - i J_1 ) (k{\mu}) ]  \cr
i \lambda [ (J_0 + i J_1 ) (k{\lambda}) \,
 J_0 (k{\mu}) ] \, + \, i \mu [ J_0 (k{\lambda})  \,
 (J_0  - i J_1 ) (k{\mu}) ]  \cr
- 2 i \sqrt{\lambda \mu } \, e^{i \phi} [ J_0 (k{\lambda})  J_1 (k{\mu})
+ J_1 (k{\lambda})  J_0 (k{\mu}) ]  \cr }
 \right.  \cr
& \qquad + \left. e^{2 i \phi} \,
\pmatrix{ - i \mu [ J_1 (k{\lambda})  \,
 (J_0 + i J_1 ) (k{\mu}) ] \, + \, i \lambda
[ (J_0  - i J_1 ) (k{\lambda}) \,  J_1
(k{\mu}) ] \,
\cr
- \mu [ J_1 (k{\lambda})  \,
 (J_0 + i J_1 ) (k{\mu}) ] \, + \, \lambda
[ (J_0  - i J_1 ) (k{\lambda}) ] \,  J_1 (k{\mu})  \,   \cr
0 \cr } \right]  . \qquad  & (A7) \cr
} $$
It is interesting to see the behaviour of this field $ {\bf F}_Z $ along the
$x_3$ axis. For the positive $x_3$ axis $ \mu = 0 $ and $ \lambda = x_3 \, ,$
and the field is just
$$ \eqalignno{
2 \big({\bf F}_Z \big)_{\mu = 0}
&=  e^{ - i k t } \, \pmatrix{  (J_0 + i J_1 ) (k{x_3})  \,  \cr
i (J_0 + i
J_1 ) (k{x_3})  \cr
0 \cr } \;
{\scriptstyle k x_3 \gg 1 \atop \hbox to 45pt{\rightarrowfill} }
\quad  e^{- i \pi / 4} \, \sqrt{ 2 \over \pi {|x_3|} }
\, \pmatrix{ 1  \cr i  \cr 0 \cr } e^{i k x_3 - i k t} \, . \cr
&  & (A8) \cr }  $$
For the negative $x_3$ axis $ \lambda = 0 $  and $ \mu = - x_3 \, ,$ and
we obtain the same expression (A8) for the
field along the negative $x_3$ axis. Along the $x_3$ axis the field ${\bf F}_Z $
appears as a right-circularly polarized plane wave propagating in the
$+x_3$
direction, with amplitude decaying by $ | x_3 |^{- 1 / 2} $ from the origin.
Away from the
$x_3$ axis the field decays more rapidly: the $x_1 , x_2$ plane
through the origin is  parametrized by
$ \lambda = \mu = \sqrt{x_1^2 + x_2^2} / 2 \, ,$ then substituting $ \lambda =
\mu = \rho \, $ into (3.7) one finds that the field  decays by $ \rho^{- 1 } $
from the origin. Thus the field ${\bf F}_Z $ is highly directional along the
axis of propagation.

When $ m = 1 \, , \, c > 1/2 \, ,$ the resulting
field becomes infinite at
spatial infinity.

\vskip 0.2 in

\noindent {\bf Appendix B.  Orthonormality of the basis functions $ U ({\bf k})
$  }  \hfil\break
\noindent First we will consider the orthogonality of two left-circularly
polarized waves, $  {U_L } ({\bf k}) \, ,$ $ \,
{U_L } ({\bf k'}) \, .$ Due to the fact that the
$ U_L ({\bf k}) \, , \,  U_L ({\bf k'}) $ are eigenstates of the angular
momentum in the direction of propagation
$ {\bf k} $ or $ {\bf k'} \, ,$ and the $ {\bf L} $
operator is Hermitian,
then
$$ \langle U_L ({\bf k}) | U_L ({\bf k'}) \rangle = 0  \qquad
\hbox{ when } \qquad
{\bf \hat{k}} \neq  {\bf \hat{k}}' \, .$$  So we
need only consider the case when the waves are propagating in the same direction
but with different frequencies, i.e.\ when $ {\bf k} = \{ 0 , 0 , k \} \, $
and $ {\bf k'} = \{ 0 , 0 , k' \} \, .$ Then recalling the $ U_L \, $ of (19) we
have the inner product
$$ \eqalignno{
\langle U_L &({\bf k}) | U_L ({\bf k'}) \rangle  \cr
&= - { 1
\over 4 } \int d^3 x \; \left[ \left( - \big(
 {\hat{x}_1 + i \hat{x}_2 \over 1 - \hat{x}_3 } \big)
  [ e^{ i k x_3 } - e^{ i k r } ] \,
- \, \big({\hat{x}_1 + i \hat{x}_2 \over 1 + \hat{x}_3 } \big)
   [ e^{ i k x_3 } - e^{ - i k r } ] \,  \right)  \;
\left( k' (\hat{x}_1 - i \hat{x}_2 ) \, e^{ - i k x_3 } \right) \right. \cr
& \qquad + \left. \left( - \big(
 {\hat{x}_1 - i \hat{x}_2 \over 1 - \hat{x}_3 } \big)
  [ e^{ - i k' x_3 } - e^{ - i k' r } ] \,
- \, \big({\hat{x}_1 - i \hat{x}_2 \over
1 - \hat{x}_3 } \big)
   [ e^{ - i k' x_3 } - e^{ i k' r } ] \,  \right) \;
\left( k (\hat{x}_1 + i \hat{x}_2 ) \, e^{ i k x_3 } \right) \right]  \cr
&=  \, { 1  \over 4 } \int d^3 x \; \left[ k' \left(  \big(
 {1 + \hat{x}_3 } \big)
  [ e^{ i (k -  k' ) x_3 }  - e^{ i (k r -  k' x_3 )} ] \,
+ \, \big({ 1 - \hat{x}_3 } \big)
 [ e^{ i (k -  k' ) x_3 } - e^{ - i k r - i k' x_3 } ] \,  \right) \right. \;
\cr
& \qquad + \left. \, k \left(  \big(
 {1 + \hat{x}_3 } \big)
  [ e^{ i (k -  k' ) x_3 }
- e^{ - i ( k' r -  k x_3 )} ] \,
+ \, \big({ 1 - \hat{x}_3 } \big)
   [ e^{ i (k -  k' ) x_3 } - e^{ i ( k' r + i k x_3 )} ] \,  \right)
\right] \cr
&= \, { 1  \over 4 } \int d^3 x \; \left[ 2 \,
( k + k' ) \, e^{ i (k -  k' ) x_3 }
\; \right. \cr
& \qquad - \, \left.  \, \big(
 {1 + \hat{x}_3 } \big)
  [ k' e^{ i (k r -  k' x_3 )}  + k e^{ - i ( k' r -  k x_3 )} ] \,
- \, \big({ 1 - \hat{x}_3 } \big)
   [ k' e^{ - i k r - i k' x_3 } + k e^{ i (k' r + i k x_3 )} ] \,  \right] .
\cr   } $$

The first term yields the delta function $\, (2 \pi )^3  \, k \;  \delta ({\bf
k} - {\bf k'}) \, .$
We go over to spherical coordinates to evaluate the
remaining terms, putting
$ \hat{x}_3  = \cos \theta = \nu \, ,$ the rest of the integral
is
$$ \eqalignno{
& -\, { 2 \pi  \over 4 } \int_0^\infty r^2 dr \int^1_{- 1} d\nu \;
\left( \big(  {1 + \nu } \big)
  [ k' \, e^{ i r (k -  k' \nu )}  + k \, e^{ - i r (k' -  k \nu )} ] \,
+ \, \big({ 1 - \nu } \big)
 [ k' \, e^{ - i r (k  +  k' \nu )} + k \,
e^{ i r (k' +  k \nu )} ] \,  \right)
\cr
&=  -\, { 4 \pi  \over 4  } \int_0^\infty r^2 dr \int^1_{- 1} d\nu \;
\left(  \big(
 {1 + \nu } \big)
  [ k' \, \cos (r (k -  k' \nu ) )  + k \, \cos (r (k' -  k \nu ) ) ] \,
\right)
\cr
&= -\, { 4 \pi  \over 4 } \int_0^\infty r^2 dr \;
(\partial_k - \partial_{k'} ) \int^1_{- 1} d\nu \;
\left(  {1 \over  r }
[ k' \, \sin (r (k -  k' \nu ) )  - k \, \sin (r (k' -  k \nu ) )  ] \,
\right)
\cr
&= -\, { 4 \pi  \over 4 } \int_0^\infty r^2 dr \;
(\partial_k - \partial_{k'} )
\bigg|^1_{- 1}  \, {1 \over r^2 }
[ \cos (r (k -  k' \nu ) )  - \cos (r (k' -  k \nu ) )  ]  \,  \bigg| \,
= \, 0 \, , \cr
} $$
so that finally
$$ \langle U_L ({\bf k}) | U_L ({\bf k'}) \rangle
=  (2 \pi )^3  \, k \;  \delta ({\bf k} - {\bf k'}) \, $$
with $ k \equiv | {\bf k} | \, .$
The other identities of (31) follow similarly.

\beginsection References

\item{$^1 $} A.C.T. Wu, {\it Debye scalar potentials for the electromagnetic
fields }
{Phys.Rev.D} {\bf 34}, 3109-3110 (1986)
\item{$^2 $} C.G. Gray, {\it Multipole expansions of electromagnetic fields
using Debye potentials }
{Am. J. Phys.} {\bf 46}, 169-179 (1978)
\item{$^3 $} See the notes of Ref. 2 for the original work on the Debye
potentials
\item{$^4 $} J.R. Reitz, F.J. Milford and
R.W. Christy, {\it Foundations of Electromagnetic Theory} (Addison - Wesley,
Reading MA, 1993)
\item{$^5 $} J.D. Jackson, {\it Classical Electrodynamics}
(John Wiley \& Sons, New York, 1975) 2nd ed.
\item{$^6 $} C.J. Bouwkamp and H.G.B. Casimir,
{\it On multipole expansions in the theory of electromagnetic radiation }
{Physica} {\bf 20}, 539-554 (1954)
\item{$^7 $} C.H. Wilcox, {\it Debye potentials } {J.Math.Mech.} {\bf 20},
167-201 (1957)

\item{$^8 $} R. Courant and D. Hilbert, {\it Methods of Mathematical
Physics, Vol 1} (Interscience Publishers Inc, New York, 1953) pp.378
\item{$^9 $} See for example pp. 767-769 of Ref. 5
\item{$^10 $}  S.N. Mosley, {\it Electromagnetics in retarded time and
photon localization } {Am.J.Phys.} {\bf 65}, 1094-1097 (1997)
\item{$^11 $} See pp. 245-251 of Ref. 5
\item{$^12 $} R.H. Good and T.J. Nelson, {\it Classical Theory of Electric
and Magnetic Fields}  (Academic Press, New York, 1971) pp.609
\item{$^13 $} S.S. Schweber, {\it An Introduction to
Relativistic Quantum
Field Theory}  (Harper and Row, New York, 1961) pp.117
\item{$^14 $} I. Bialynicki-Birula {\it The photon wavefunction } in
{\it Progress in Optics Vol 36 } (Ed. E. Wolf, Elsevier, Amsterdam, 1996) pp.
246-294
\item{$^15 $} I. Bialynicki-Birula and Z. Bialynicka-Birula, {\it Quantum
Electrodynamics}  (Pergamon Press, Oxford, 1975) pp.142
\item{$^16 $} L.D. Landau and E.M. Lifshitz, {\it The Classical Theory of
Fields}  (Pergamon
Press, Oxford, 1971) 3rd revised ed., pp. 119-123
\item{$^17 $} Any complex vector $ {\bf C}  $ can be split into three
components with the following set of projection operators:
$ \, {\bf P}_{\uparrow} \equiv {\scriptstyle{1 \over 2}} (1 - {\bf \hat{x}} \,
{\bf \hat{x}} \cdot  + i \, {\bf \hat{x}} \, \times ) \, , \;
{\bf P}_0 \equiv  {\bf \hat{x}} \, ({\bf \hat{x}} \cdot \, ) \, , \;
{\bf P}_{\downarrow} \equiv {\scriptstyle{1 \over 2}} (1 - {\bf \hat{x}} \,
{\bf \hat{x}} \cdot -
i \, {\bf \hat{x}} \, \times ) \, ,$
satisfying
$ {\bf P}_{\epsilon} {\bf P}_{\epsilon'} = \delta_{\epsilon \epsilon'} \,
{\bf P}_{\epsilon} \, ,$ with $ \epsilon = \{ {\uparrow} , 0 , {\downarrow}
\} \, .$ The transverse components $ ({\bf P}_{\uparrow}{\bf C}) ,
({\bf P}_{\downarrow}{\bf C})  $ can then each be
expressed in terms of a complex potential by
$ ({\bf P}_{\uparrow}{\bf C}) = {\bx \nabla}_{\uparrow} \chi
\equiv {\bf P}_{\uparrow} {\bx \nabla} \chi \, $ and $ \,
({\bf
P}_{\downarrow}{\bf C}) = {\bx \nabla}_{\downarrow} \zeta
\equiv {\bf P}_{\downarrow} {\bx \nabla} \zeta \, .$
See the paper by Wilcox (Ref.7) for equivalent methods of expressing a
transverse vector
$ ({\bf P}_{\uparrow}{\bf C} + {\bf P}_{\downarrow}{\bf C}) $ in terms of two
potentials.
\item{$^18 $} R. Resnick, {\it Introduction to Special Relativity}
(John Wiley \& Sons, New York, 1968) pp.169
\item{$^19 $} D.K. Monroe, {\it Lorentz transformation of Debye
potentials
}  {J.Math.Phys.} {\bf 25}, 1787-1790 (1984)
\item{$^20 $} E. Witten, {\it Duality, spacetime and quantum mechanics }
{Phys.Today} {\bf 50}(5), 28-33 (1997)
\item{$^21 $} H. Buchholz, {\it The Confluent Hypergeometric
Function}  (Springer-Verlag, Berlin, 1969) pp.50
\item{$^22 $} H. Hochstadt, {\it The Functions of Mathematical Physics }
(Wiley Interscience, New York, 1971) pp.192

\bye